\documentclass[useAMS,usenatbib]{mn2e}
\usepackage{graphicx}
\usepackage{color}

\title[Eccentricities of Transiting Planets]{Determining Eccentricities of Transiting Planets:\\ A Divide in the Mass-Period Plane}

\begin{document}
\bibliographystyle{mn2e}

\author[F. Pont et al.]{Fr\'ed\'eric Pont$^1$, Nawal Husnoo$^1$,   Tsevi Mazeh$^2$, Daniel Fabrycky$^{3}$\\
$^1$ School of Physics, University of Exeter, Exeter, EX4 4QL, UK \\
$^2$ School of Physics and Astronomy, Tel Aviv University, Tel Aviv 69978, Israel\\
$^3$ Harvard-Smithsonian Centre for Astrophysics, Garden Street, Cambridge MA}


\maketitle

\label{firstpage}

\begin{abstract}
The two dominant features in the distribution of orbital parameters for close-in exoplanets are the prevalence of circular orbits for very short periods, and the observation that planets on closer orbits tend to be heavier. The first  feature is interpreted as a signature of tidal evolution, while the origin of the second, a ``mass-period relation'' for hot Jupiters, is not understood.  In this paper we re-consider the ensemble properties of transiting exoplanets with well-measured parameters, focussing on orbital eccentricity and the mass-period relation. We recalculate the constraints on eccentricity in a homogeneous way, using new radial-velocity data, with particular attention to statistical biases. We find that planets on circular orbits gather in a well-defined region of the mass-period plane, close to the minimum period for any given mass. Exceptions to this pattern reported in the Literature can be attributed to statistical biases.  The ensemble data is compatible with classical tide theory with orbital circularisation caused by tides raised on the planet, and suggest that tidal circularisation and the stopping mechanisms for close-in planets are closely related to each other.  The position mass-period relation is compatible with a relation between a planet's Hill radius and its present orbit.







\end{abstract}



\section{Introduction}

Transiting planets are presently our main source of information on the formation, structure and evolution of extra-solar planets\footnote{see compilations in www.exoplanets.eu (unedited catalogue of published parameters), www.exoplanets.org (edited catalogue focussing on Doppler detections), www.inscience.ch/transits (edited catalogue of transiting planets), and in \citet{sou09}.}. Thanks to the success of photometric transit surveys, enough transiting planets are now known to start applying more robust statistics to their ensemble features. In particular, transiting planets help to address the question of two puzzling features of the presently known population of exoplanets: the presence of planets on very close orbits, and the wide range of orbital eccentricities. 

Because of the strong selection biases favouring shorter orbital periods, most known transiting planets have close orbits ($a<0.1$ AU). Contrary to the sample of planets found by radial-velocity surveys at larger orbital distances, the orbits of transiting planets tend to be more circular, a tendency that is interpreted as a signature of tidal circularization \citep{maz08}. A close study of the eccentricity distribution of these objects needs to address two related issues: how did close-in planets get there, and what is the role of tidal orbital evolution? Early inward migration by gravitational interaction with the protoplanetary gas disc is the favoured explanation for the presence of close-in planets \citep{lin96}, with planet-planet scattering, followed by tidal dissipation, as another possibility \citep{ras96}.

The empirical evidence that has to be accounted for by a successful understanding includes the pile-up of close-in planets near orbital periods of  3 days, the predominance of circular orbits inwards of a few days, and the possible existence of a correlation between mass and period for close-in gas giants (with more massive hot Jupiters found closer in, $P$=1--2 days rather than 3--4 days), as pointed out by  \citet{maz05}.

The classical theory of tides predicts that the orbits of transiting planets will be circularised by the tidal interaction induced by the star on the planet, with a timescale increasing sharply with orbital distance \citep[e.g.][]{gold66}. The observed transition from eccentric to circular orbits around periods of a few days is qualitatively compatible with this prediction. Similar behaviour has been found for binaries, e.g. \citet{mat88} and \citet{maz08}. 

\begin{figure}
\resizebox{\hsize}{!}{\includegraphics{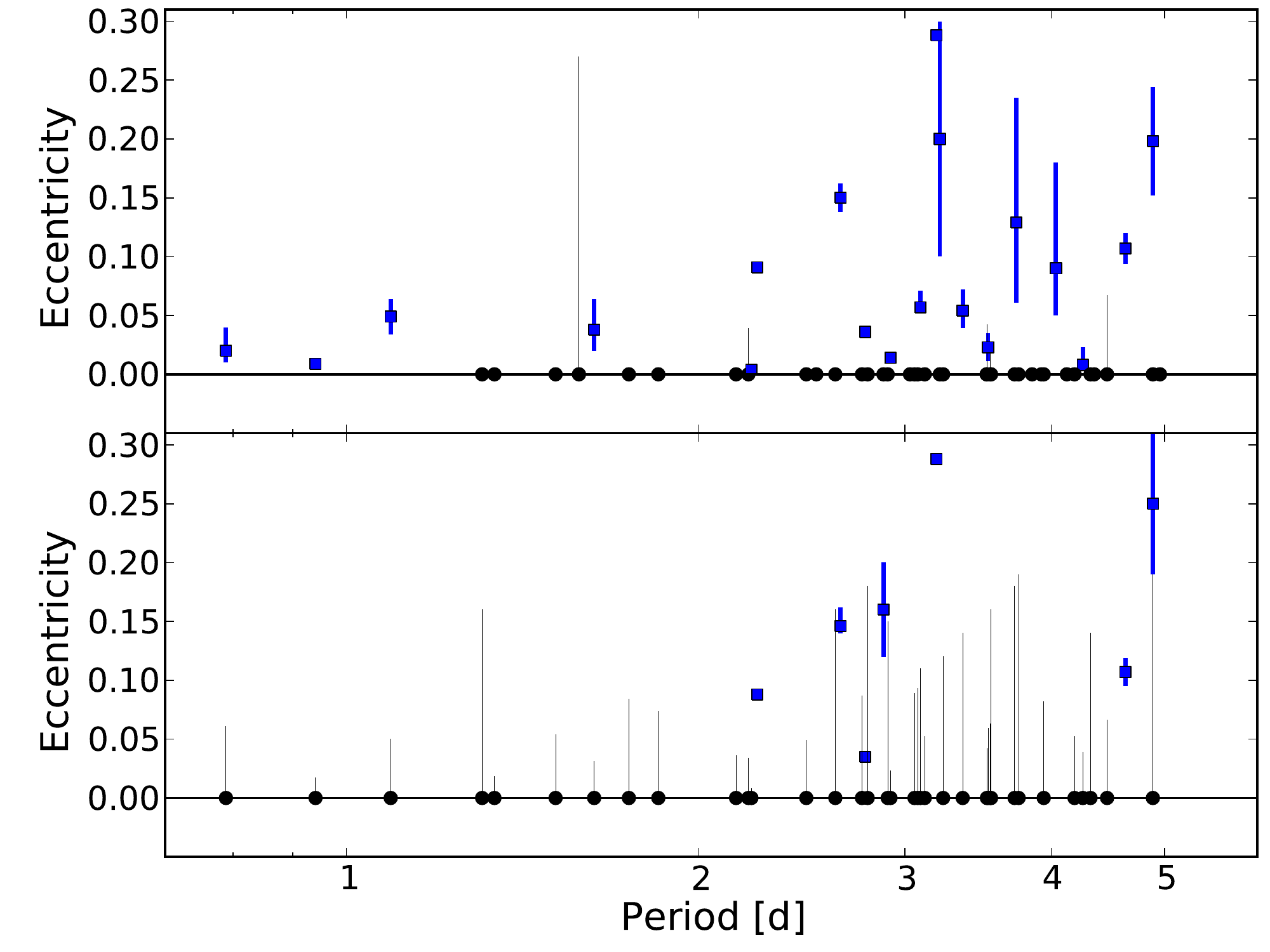}}
\caption{Period-eccentricity plot for transiting planets using a compilation of the latest values published in the literature {\bf (top)} compared to the determinations in the present study {\bf (bottom)}. The 95\% lower intervals are shown when a null eccentricity is not excluded at the 5\% level. Note the difference for low-eccentricity, short-period values.}
\label{exoe}
\end{figure}

Recently though,  a number of small but non-zero orbital eccentricities have been reported for several planets on extremely close orbits (below 3 days), including 
WASP-12b \citep{heb09}, WASP-14b \citep{jos09}, WASP-18b  \citep{hel09}, WASP-19b  \citep{heb10} and GJ 436b \citep{but04} --- see top panel of Fig.~\ref{exoe}. Specific scenarios have been invoked to explain these apparently anomalous cases, such as the presence of a perturbing companion \citep{rib08,mar08}, or widely different coefficients in the intrinsic response of planets to tides \citep{mats09}. Another relevant recent development is the detection of several systems with strong spin-orbit misalignment \citep[][and references therein]{win10}, which is presently interpreted as an indication that the disc-migration scenario is in need of significant updates, and that dynamical evolution after the dissipation of the disc is probably more important than previously thought, and even possibly dominant in determining the orbital properties of close-in planets.

We have been performing a series of observational programmes to gather more radial-velocity (RV) data on known transiting planets, in order to study these issues. In this paper, we revisit the data on orbital eccentricity for transiting planets, based on new data and a re-analysis of extant data, and examine the implications in terms of the ensemble properties of close-in planets regarding the orbital circularisation and the stopping mechanism.

\section{Planet sample and orbit analysis}

Our sample consists of all known transiting planets fulfilling the following selection criteria: (i) objects with precisely determined parameters (i.e. planetary radius and planetary mass measured to better than 10\% accuracy) (ii) host stars brighter than $V$=15 mag (iii) objects known from peer-reviewed publication by 1st July 2010.

We use published radial velocity data as well as new HARPS data from our observation programme for WASP-2, WASP-4, WASP-5 and WASP-7.  The radial velocity data for these objects is presented and discussed in more details in \citet{hus11}, and is available as an electronic table. 

The brightest known transiting planets were identified by Doppler planet surveys first and only subsequently found to be transiting. Their orbital eccentricity is therefore usually well determined, because a sizeable number of individual radial velocity measurements is required to find the period of a planetary orbit. 
Most of our sample, however, consists of objects discovered by photometric searches for transits. In that case, the number of radial velocity measurement is lower, since the RV data are only required to detect the presence of an orbital signal at the period and phase of the transits. Transiting planets from photometric searches are on average much fainter than the targets of Doppler searches,  so that in general the RV data is also less precise. In these case,  there are three options to calculate the orbital parameters:
\ \\
- the eccentricity is set to zero, under the assumption that it is small and undetectable, and the data is solved for the other parameters of the system (e.g. the planets from the OGLE, HAT and TrES surveys)
\ \\
- the RV data clearly indicate an eccentric orbit, and the eccentricity $e$ is solved together with the other parameters (e.g. HAT-P-2).
\ \\
- a Monte-Carlo Markov Chain (MCMC) integration is run on all data, with the eccentricity as a free parameters (e.g. planets from the WASP survey).

The problem with the third method is that there is an intrinsic bias in the determination of eccentricity from radial-velocity data, because measurement errors on RV data for a circular orbit will always give rise to a finite best-fit eccentricity. This effect was studied in the context of stellar binaries by \citet{luc71}  and more recently by \citet{Shen2008}. As a result, even when the measurement uncertainties are correctly estimated, the centre of the posterior distribution from a MCMC for a circular orbit will tend to be 1-3 sigma away from zero. An additional issue is the presence of other sources of signal in the radial velocity data. The most common causes are stellar activity, instrumental drifts, and additional companions in the system. When these sources are not included in the orbital analysis, they may induce spurious eccentricity detections  \citep{Rodigas2009,hus10}  because any radial-velocity offset from a circular orbit will make it appear more eccentric. 




A large fraction of published eccentricities for transiting planets are significantly different from zero at the few-sigma level only. Particular attention to this issue is justified to avoid spurious detections, which would be especially confusing in the context of studying tidal evolution, since they would tag some circularised objects as having low but significant eccentricity, thus being good candidates for rapid on-going tidal evolution. 


The recent case of WASP-12, discussed in \citet{heb09,Morales2009,Cam10,hus10}, shows how false positives can arise in non-circular orbit detections. The orbital eccentricity of WASP-12b, initially thought to be significant at the 3-sigma level, was subsequently found to be spurious by further RV measurements and the detection of the secondary eclipse by the Spitzer satellite. Another example is WASP-10, recently re-analysed by \citet{mac10}. These authors conclude that the initially significant eccentricity detection was in reality due to radial-velocity variations induced by stellar activity. 


For these reasons, we re-analyse the RV data for transiting planets with a specific attention to the discrimination between circular and eccentric orbits. 

We use a Markov Chain Monte Carlo (MCMC) method with the Metropolis-Hastings algorithm \citep[see ][for example]{Cameron2007a}. Our implementation is described in \cite{Pont2009b}. 
We take into account the possible non-random RV uncertainties in addition to the formal error bars, from either physical or instrumental causes, ensuring that the size of the $O-C$ residuals is compatible with the uncertainties used for the MCMC. We include the possibility of a certain amount of correlation in the additional RV noise \citep[using the methods in ][]{pon06}. Finally, to decide whether the evidence for an eccentric orbit is statistically significant,  we examine the ``$e=0$'' possibility in hypothesis-testing mode rather than in parameter-estimation mode. In other words, we do not simply adopt the best-fitting value of $e$ as the most likely value, but also calculate if a non-zero value of $e$ is more likely than the $e=0$ null hypothesis given the addition of free parameters and the possibility of non-random noise in the data.  We do this by calculating the BIC (Bayesian Information Criterion) of the best-fit eccentric orbit compared to the best-fit circular orbit. An eccentric Keplerian has two extra free parameters compared to a circular one, so that a closer fit is always expected even in the absence of real evidence. This last step penalises the eccentric solution for the two extra parameters.

In summary, we consider the evidence for orbital eccentricity to be sufficient not only when an eccentric orbit gives a better fit to the data at a level higher than the uncertainties, but when the BIC value shows that the improvement compared to a circular orbit is significantly larger than would be expected from instrumental and physical sources of noise and added free parameters.

To circumvent the difficulty posed by the tight correlation of the $e$ and $\omega$ parameters near $e=0$, we use $e\cos\omega$ and $e\sin\omega$ as parameters instead of $e$ and $\omega$, following \citet{Ford2006}. Using these parameters also makes it easier to avoid biasing results towards higher eccentricities when expressing the output of the MCMC chain in terms of confidence intervals. The posterior probability distribution for $e$ is one-sided near $0$ because $e<0$ is forbidden, but  $e\cos\omega$ and $e\sin\omega$ do not have this problem.  We calculate the central values of the probability distribution of $e$ and $\omega$ from the central values for $e\cos\omega$ and $e\sin\omega$ in the output of the MCMC.

\section{Results: orbital eccentricities for transiting planets}


Table~\ref{table1} shows the new eccentricity determinations. When the detection of eccentricity is significant, the value determined with the MCMC fit for the eccentric orbit is given, together with the 68\% central confidence interval. For systems where our analysis showed there was no evidence for an eccentric orbit, the upper 95\% confidence interval on the eccentricity is given.



\begin{table*}

\centering
\begin{tabular}{l l r l l } \hline
Name & Mass& Period & Eccentricity  & Eccentricity  \\ 
&  [M$_j$] & [days] & (this study)  & (literature)  \\ \hline
CoRoT-1b 			& 1.03 $\pm$ 0.12 & 1.509 	&    $<$   0.054   & 0 \\
CoRoT-2b 			& 3.31 $\pm$ 0.16 &  1.743	&  $<$ 0.084     & 0 \\
CoRoT-3b			& 21.2 $\pm$ 0.8 &  4.257	& $<$ 0.039 & $0.008^{+0.015}_{-0.005}$ \\
CoRoT-4b 			& 0.72 $\pm$ 0.08	&  9.202	&    $<$   0.39    &  0$\pm$0.1 \\
CoRoT-5b 			& 0.47 $\pm$ 0.04 &  4.038	&   $<$   0.23  & 0.09$^{+0.09}_{-0.04}$ \\
CoRoT-6b 			& 2.96 $\pm$ 0.34 &  8.887  	&  $<$   0.41    & $<$0.1\\
CoRoT-9b 			& 0.84 $\pm$ 0.07 &  95.274	&  0.13 $\pm 0.04$  &   0.11  $\pm 0.04$  \\
CoRoT-10b			& 2.75 $\pm$ 0.16 &  13.241	& 0.53 $\pm 0.04$  $^{*}$ & 0.53$\pm$0.04 \\
GJ-436b 				& 0.074 $\pm$ 0.005 &  2.644	& 0.146 $^{+0.016} _{-0.006}$  & $0.150\pm0.012$ \\
GJ-1214b 			& 0.021 $\pm$ 0.003 &  1.580	&   $<$   0.45  & $<0.27 (95\%)$\\
HAT-P-1b 			& 0.52 $\pm$ 0.03 &  4.465	& $<$   0.066   & $<$0.067 (99$\%$) \\
HAT-P-3b 			& 0.60 $\pm$ 0.03 &  2.900	& $<$   0.15   & 0 \\
HAT-P-4b 			& 0.68 $\pm$ 0.04 &  3.056  	& $<$   0.087    & 0 \\
HAT-P-5b 			& 1.06  $\pm$ 0.11 &  2.788  	& $<$   0.18   & 0 \\
HAT-P-6b 			& 1.06 $\pm$ 0.12 &  3.853 	& $<$   0.35    & 0 \\
HAT-P-7b 			& 1.82  $\pm$ 0.03 &  2.205	&$<$   0.044    & $<$0.039 (99\%) \\
HAT-P-8b 			& 1.52 $\pm$ 0.18 &  3.076	&     $<$   0.10   & 0 \\
HAT-P-9b 			& 0.78 $\pm$ 0.09 &  3.923	&     $<$   0.40  & 0 \\
HAT-P-10b/WASP-11b 	& 0.46 $\pm$ 0.03 &  3.722	&     $<$   0.18     & 0 \\
HAT-P-11b 			& 0.08 $\pm$ 0.01 &  4.888	&  0.23 $\pm 0.06$  & 0.198$\pm$0.046 \\
HAT-P-12b 			& 0.21 $\pm$ 0.01 &  3.213	&     $<$   0.22  & 0 \\
HAT-P-13b 			& 0.85 $\pm$ 0.04 &  2.916	&     $<$   0.023   $^{*}$& 0.014$\pm$0.005 \\
HAT-P-14b 			& 2.23 $\pm$ 0.06 &  4.628	&  0.107 $\pm$0.012  & 0.107$\pm$0.013 \\
HAT-P-15b 			& 1.95 $\pm$ 0.07 &  10.864	&  0.19 $\pm 0.02$ $^{*}$ & 0.190$\pm$0.019 \\
HAT-P-16b 			& 4.19 $\pm$ 0.09 &  2.776	&  0.035 $\pm 0.003$  &  0.036$\pm$0.004\\
HD17156b 			& 3.21 $\pm$ 0.08 &  21.217	&  0.684 $\pm 0.002$ $^{*}$& 0.6835$\pm$0.0017 \\
HD80606b 			& 3.94 $\pm$ 0.11	&  111.436	& 0.934  $\pm 0.001$  $^{*}$ & 0.93366$_{-0.00043}^{+0.00014}$ \\
HD147506b/HAT-P-2b 	& 8.74 $\pm$ 0.25 &  5.633	&  0.514 $\pm 0.006$  & 0.517$\pm$0.003 \\
HD149026b		 	& 0.356 $\pm$ 0.013 & 2.877 & 0.16$\pm$0.04 & 0\\
HD189733b 			& 1.15 $\pm$ 0.03 &  2.219	&     $<$ 0.008 $^{*}$  & 0.004$\pm$0.003 \\
HD197286b/WASP-7b 	& 0.96 $\pm$ 0.18 &  4.955	&     $<$   0.25   & 0 \\
HD209458b 			& 0.64 $\pm$ 0.09 &  3.525	&     $<$   0.042  $^{*}$ & $<$   0.042 \\
Kepler-4b 			& 0.077 $\pm$ 0.012 &  3.213	&     $<$   0.35 $^{*}$  & 0.2$\pm$0.1 \\
Kepler-5b 			& 2.11 $\pm$ 0.06 &  3.548	&     $<$   0.063  $^{*}$  & $<$0.024 \\
Kepler-6b 			& 0.67 $\pm$ 0.03 &  3.235	&     $<$   0.12    & 0\\
Kepler-7b 			& 0.43 $\pm$ 0.041+0.04 &  4.886	&     $<$   0.19   &  0\\
Kepler-8b 			& 0.60 $\pm$ 0.19 &  3.523 	&     $<$   0.64   & 0 \\
TrES-1b				& 0.76 $\pm$ 0.05 &  3.030	&     $<$   0.24    & $0$ \\
TrES-2b 				& 1.25 $\pm$ 0.05 &  2.471	&     $<$   0.098    &  0 \\
TrES-3b 				& 1.91 $\pm$ 0.08 &  1.306	&     $<$   0.16    & 0 \\
TrES-4b 				& 0.88 $\pm$ 0.07 &  3.554	&     $<$   0.16   & 0 \\
WASP-1b 			& 0.86 $\pm$ 0.07 &  2.520	&     $<$   0.28    & 0 \\
WASP-2b 			& 0.847 $\pm$ 0.045	&  2.152	&     $<$   0.036   & 0 \\
WASP-3b 			& 2.07 $\pm$ 0.07 &  1.847	&     $<$   0.068    & 0 \\
WASP-4b 			& 1.24 $\pm$ 0.07 &  1.338	&     $<$   0.018    & 0 \\
WASP-5b 			& 1.64 $\pm$ 0.08	&  1.628	&     $<$   0.031   & 0.038$^{+0.026}_{-0.018}$ \\
WASP-6b 			& 0.50 $\pm$ 0.04 &  3.361	&     $<$   0.14    & 0.054$^{+0.018}_{-0.015}$ \\
WASP-10b 			& 3.0 $\pm$ 0.2 &  3.093	&     $<$   0.11 & 0.057$^{+0.014}_{-0.004} $\\
WASP-12b 			& 1.4 $\pm$ 0.1 &  1.091	&     $<$   0.050    & 0.049$\pm$0.015 \\
WASP-13b 			& 0.46 $\pm$ 0.05 &  4.353	&     $<$   0.36    & 0 \\
WASP-14b 			& 7.3 $\pm$ 0.5 &  2.244	&       0.088 $\pm 0.003$    & 0.091$\pm$0.004 \\
WASP-15b 			& 0.54 $\pm$ 0.05 &  3.752	&     $<$   0.19   & 0 \\
WASP-16b 			& 0.86 $\pm$ 0.08 &  3.119	&     $<$   0.052  & 0 \\
WASP-17b 			& 0.49 $\pm$ 0.06 &  3.735	&     $<$   0.31   & $0.129^{+0.106}_{-0.068}$ \\
WASP-18b 			& 10.4 $\pm$ 0.4 &  0.941	&     $<$   0.017    & 0.009$\pm$0.003 \\
WASP-19b 			& 1.15 $\pm$ 0.08 &  0.789	&     $<$   0.061    & 0.02$\pm$0.01 \\
WASP-21b 			& 0.30 $\pm$ 0.01	&  4.322	&    $<$   0.14    & 0 \\
WASP-22b 			& 0.56 $\pm$ 0.02 &  3.533	&     $<$   0.059  & 0.023$\pm$0.012 \\
WASP-26b 			& 1.02 $\pm$ 0.03 &  2.757	&     $<$   0.087   & 0 \\
XO-1b 				& 0.90 $\pm$ 0.07 &  3.942	&     $<$   0.29    & 0 \\
XO-2b 				& 0.57 $\pm$ 0.06 &  2.616	&     $<$   0.41  & 0 \\
XO-3b 				& 11.8 $\pm$ 0.6 &  3.192	&  0.288 $\pm 0.004$  $^{*}$& 0.2884$\pm$0.0035 \\
XO-4b 				& 1.7 $\pm$ 0.2 &  4.125	&     $<$   0.61 &  0\\
XO-5b 				& 1.06 $\pm$ 0.03 	&  4.188	&  $<$ 0.052  &  0\\ \hline
\end{tabular}
\caption{Eccentricity of transiting planets. The list includes planets with well-determined parameters as of July 2010. When a circular orbit is compatible with the data according to our analysis, we give the upper 95\% confidence limit. Asterisks mark values that we adopted from the literature without re-analysis. An up-to-date list of parameters is kept on http://www.inscience.ch/transits. }
\label{table1}
\label{table2}
\end{table*}


Fig.~\ref{exoe} compares our eccentricity determinations with a compilation of published values. Some objects were already analysed with methods similar to our own, and in that case we adopt the eccentricity distribution from the Literature  \citep{lau05,boi09,nar09,win09d,bon10,kip10,kov10,win10}. Compared to published values, our new RV data and re-analysis result in better constraints of the eccentricity for several objects, and significant changes for others. But the main result is that in several cases, an eccentricity detection at the few-sigma level is transformed into an upper limit, i.e. the evidence for orbital eccentricity is found to be insignificant.  Specifically, there are 10 transiting hot Jupiters with eccentricity detections significant at more than the 2-sigma level below period of 5 days in the literature. In our re-analysis, there are five below 5 days, none of which are in the 0.5-2 $M_J$ mass range typical of hot Jupiters. The systems with $>$2$\sigma$ published eccentricity detections that we do not find to be significant are CoRoT-5, HAT-P-13, WASP-5, WASP-6, WASP-10, WASP-12, WASP-17, WASP-18.

There are two main causes for this: either new RV measurements show the initial eccentricity detection to have been spurious, or the hypothesis-testing statistic shows that the difference between the best-fitting eccentric orbit and a circular solution is not significantly larger than would be expected from intrinsic bias and measurements errors, and therefore a circular orbit is compatible with the data. With our new RV data, we have found evidence of RV modulations caused by stellar activity, instrumental effects or an additional companion in the system for WASP-4 and WASP-5. For objects measured with the SOPHIE spectrograph in the ``HE'' mode (used for fainter objects), we also find large long-term instrumental drifts, that need to be included in the error budget. 


\definecolor{gray}{rgb}{0.7,0.7,0.7}

\begin{figure*}
\resizebox{\hsize}{!}{\includegraphics{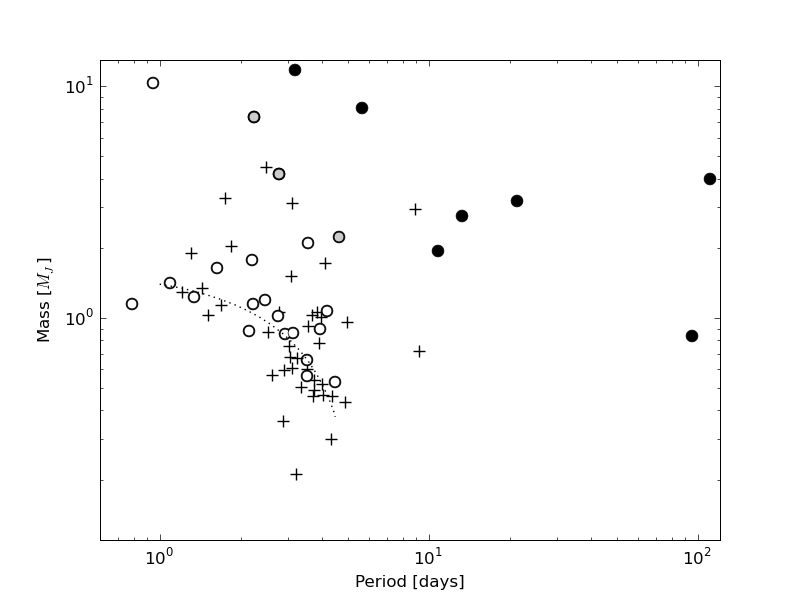}}
\caption{Mass-period diagram for transiting planets. Open circles ({\large $\circ$}) indicate objects with well-constrained orbital eccentricity compatible with zero, filled circles  ({\large $\bullet$}) show definitely eccentric orbits with $e>0.1$, and crosses  ({+}) objects for which no definite conclusion could be derived from the present data. Gray circles  ({\color{gray} \large $\bullet$})  indicate objects with a detected non-zero but small eccentricity (compatible with $e<0.1$, values that could remain undetected at lower masses. The dotted line is the relation from \citet{maz05}.  }
\label{MP}
\end{figure*}


The most striking feature of Table~\ref{table1} is the scarcity of confirmed eccentricity detections below the 10\% level compared to published values (see Fig.~\ref{exoe}). WASP-14, with $e=0.088 \pm 0.003$ and $P=2.2$ days, becomes the shortest-period orbit with significant eccentricity detection, and HAT-P-16 ($e=0.035 \pm 0.003$) the smallest detected eccentricity. 

Note that the scarcity of confirmed $e<0.1$ orbits is not simply a reflection of the fact that smaller eccentricities are more difficult to detect. For several objects, the uncertainty on $e$ is smaller than 1 percent, which would have allowed the detection of a small but non-zero eccentricity (this is the case either for bright and well-measured objects like HD 189733,
 or for objects with Spitzer secondary eclipse measurements like WASP-18).

For lower-mass planets (below $0.5 M_J$) the eccentricity measurement gets difficult because the RV amplitude is smaller, and the constraints on $e$ are rarely good enough to be significant in our context. That is a bias to keep in mind when studying the sample as a whole. We therefore need to compare significant eccentricities with eccentricities compatible with zero only when the uncertainties are small enough that a significant eccentricity would have been detected. 


\section{Discussion: Tidal circularisation and the mass-period relation}



%

Figure~\ref{MP} displays our sample in the mass-period diagram. The different symbols indicate probably circular orbits, eccentric orbits, and unsolved cases. In the ``probably circular'' category, we include all objects with RV data compatible with a circular orbits and excluding an eccentricity larger than 0.10 at more than the 3-sigma level. 

Figure~\ref{MP}  shows a clear relation between orbital eccentricity and the position in the mass-period diagram, much clearer than is the case when a compilation of published values of eccentricities is used. In fact, there is no confirmed eccentric orbit at all in the main ``hot Jupiter clump'' (at $P$=1--4 days and $M$=0.5--2 ${\rm M_J}$). A contrario, all objects that are on wider orbits and/or are more massive have eccentric orbits.

Another conclusion is that mass and period are correlated for the objects with circular orbits --- if not by a tight relation as proposed by \citet{maz05}, then definitely a trend, with heavier planets closer in. Especially remarkable is the tight grouping of lighter ($M< M_J$) planets on circular orbits.

These two features are our main results: close-in planets belong to two clear classes in regard to orbital parameters -- either they follow circular orbits and a mass-period trend, or they have eccentric orbit and are off the mass-period trend (on the side of larger period at a given mass). This is very strongly suggestive of tidal effects.



\begin{figure*}
\resizebox{\hsize}{!}{\includegraphics{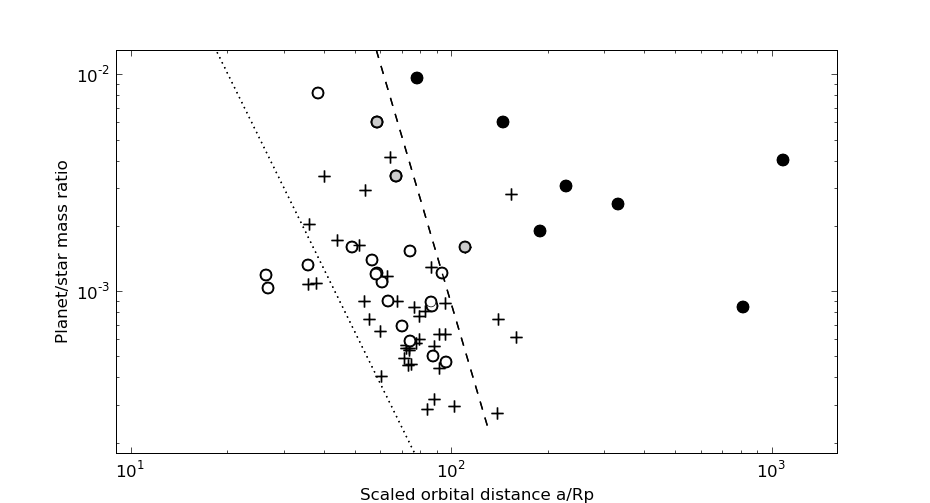}}
\caption{``Tidal'' diagram for transiting planets, showing the mass ratio of the system $M_p/M_*$ versus orbital distance in terms of planetary radius ($a/r_p$). Same symbols as on Fig.~\ref{MP}. 
The dashed line shows a 1 Gyr circularization timescale isochrone, using $Q=10^6$, $P=3$ days and $e=0$. The dotted line shows $a=2 \cdot a_H$, the location predicted by tidal circularisation from very eccentric orbits (see text).
}
\label{tidal}
\end{figure*}

%



Theoretical treatments of tidal evolution predict that the timescale of tidal circularisation will decrease with the planet-to-star mass ratio, and increase steeply with the semi-major axis scaled to the planetary radius. For the orbital circularisation of planetary orbit, the relevant effect is the tide raised on the planet by the star rather than the opposite, see e.g. Hansen et al. 2010 for a quantitative discussion.  \citet{gold66} obtain $\tau_e = \frac{4}{63} Q \left( \frac{a^3}{GM_*}\right)^{1/2}\frac{M_p}{M_*}\left(\frac{a}{R_p}\right)^5$ for the circularisation timescale, where $Q$ is the tidal quality factor --- a parametrisation of the response of the planet's interior to tidal perturbation, $\tau_e$ is the circularisation timescale, $G$ the gravitational constant and $a$ the orbital semi-major axis. When the objects in our sample are ranked by $\tau_e/Q$, we find that circular and eccentric orbits are separated as expected from classical tide theory, with an overlap in $\tau_e/Q$ between the smallest measured eccentricities and the circular orbits over less than one decade, that can be accounted for by expected variations in $Q$ and age from object to object and do not require additional factors like perturbing companions or anomalous tidal evolution.

Figure~\ref{tidal} shows the position of our sample in a plane more closely related to the circularisation timescale than the mass-period diagram. Figure~\ref{tidal} displays the planet-to-star mass ratio $M_p/M_*$, as a function of the orbital distance scaled to the planetary radius $a/R_p$. If $R_p$ and $M_s$ are much less variable than $M_p$ and $P$, and $P$ and $a$ are closely related, as is the case for hot Jupiters, this plot will be similar to the mass-period plot, except that the tidal circularisation isochrones will be straight lines. The dashed line shows a 1 Gyr circularisation isochrone using $Q$=$10^6$ (a plausible value for Jovian planets, \citet{wu05}), $P=3$ days and $e\sim0$ (the dependence of the circularisation timescale on $P$ and $e$ is weak on such a log-log plot\footnote{ The timescale depends on $e$ because an eccentric orbit has a lower angular momentum than a circular one at a given period.}).  On this plot, the relation between circularization timescale and measured eccentricity is apparent. Circular and eccentric orbits are well separated and the separation is compatible with circularisation on a timescale corresponding to the age of the systems. The small but non-zero eccentricities are all in the intermediate range. Thus, the whole sample is compatible with the operation of tidal circularisation with less than one decade of variation in system ages and values of $Q$, without any exception. This is very remarkable given the variety of  planets and host stars considered. The slope of the relation shows that orbital circularisation is caused by tides raised on the planet, rather than tides raised on the star that would predict a mass dependence in the opposite direction.

The locus of the separation between circular and eccentric orbits is not compatible with timescales comparable to the lifetimes of protoplanetary discs ($\leq 10 Myr$). This suggests that the stopping mechanism for inward migration is not related to disc migration, or a coincidence would be required to explain the correspondance of circular orbits and the mass-period relation.

Another  noteworthy element is the scarcity of high-mass planets on circular orbits \citep{ras296,jack09,win10}. Although a larger sample would be necessary for definite conclusions, it seems that the response to orbital circularisation has a strong dependence on the mass of the planet: $M\!\!<\!\!1 M_J$ planets stop at the circularisation radius, $M\!\! \sim\!\! 1 M_J$ planets can  move closer to the host star, and $M\!\!> \!\!1 M_J$ planets do not reach circularisation. A possible explanation for this behaviour is that heavier planets could raise tides in the star strong enough for angular momentum exchange and orbital decay \citep{ras96, 2003S, jack09, pont2009, win10}.  Another possibility is that heavier and lighter gas giants form by different mechanisms \citep{rib07}.

\citet{fr06} pointed out that if the orbit of close-in planets evolve from initially very eccentric orbits due to interactions with other planets, the final orbital distance would be approximately twice the distance at which the planet fills its Roche lobe ($a \simeq 2 \cdot a_H$, with $a_H$ defined as the distance at which $R_P=R_H$,  where $R_H$ is the Hill radius). Using a Bayesian analysis of the sample of radial-velocity (i.e. non-transiting) planets, \citet{fr06}  found that the data was compatible with the closest planets following  $a = \alpha \cdot a_H$, the $\sin i$ uncertainty on the mass of non-transiting planets preventing a determination of the numerical value of the multiplying factor $\alpha$. 

On Fig.~\ref{tidal}, the dotted line shows the locus of $a=2 \cdot a_H$. Clearly, this relation does not constitute a good description of the present position of close-in planets on circular orbits. However, the slope of this relation seems to be a better description of the observed locus of circular orbits than the slope of the circularisation isochrone, with $\alpha \simeq 3-4 $. One possibility to explain the larger proportionality factor in the relation between circularisation distance and $a_H$ is that planets had much larger sizes earlier on.

Intriguingly, the dependence on $a_H$ seems to hold for hot-Neptune and super-Earth planets. These planets do not at all follow the mass-period relation of hot Jupiters, with the three presently known transiting super-Earth candidates all having extremely close orbits ($P<1$ days), but once scaled to the planetary radius, their position align with a similar multiple of $a_H$. The data at this point are not sufficient for definite conclusion, and this can be tested as more small transiting planets are found and characterised.

\section{Conclusion and prospects}

The empirical features above must be accounted for by any successful theory on the nature and history of hot Jupiters. The  smooth relation between eccentricity and tidal timescale, as well as the pile-up of circular orbits at specific orbital distances at circularisation, strongly suggests that tidal circularisation and the stopping mechanism of close-in planets are related. The mass-period trend also suggests that heavier planets get circularised closer to the host stars, if at all. The lack of higher-mass planets on  circularised orbits suggests that these planets end up in the star instead, possibly because they provoke tides on the star that are strong enough to transfer angular momentum from the planet's orbit to the stellar spin. 

The observed relation between orbital distance and intrinsic parameters (size and mass) for planets on circular orbits shares some features with the predictions of \citet{fr06} for inward migration caused by planet-planet scattering.

The close association of migration and circularisation suggest that the two mechanisms operate on similar timescales. If migration was much more rapid, tidal evolution would not be efficient as a stopping mechanism. This argues in favour of secular dynamical interactions to explain the presence of close-in planets rather than migration in a protoplanetary disc.

These are exciting times in the study of close-in exoplanets. As new objects are discovered by the dozen, the statistical inferences  become more robust. Many features of the sample as a whole have become clearer: the abundance of misaligned systems and their relation with the nature of the primary, the role of inward migration, the relation between the energy budget and the planet size. 
These features seem to be related by an over-arching theme: most of the  properties of close-in planets are actually controlled by their parent stars. When planets get closer than a few days in period, the star takes over their orbital properties, their atmospheric dynamics,  and probably their size and lifetime as well. 

\section*{Acknowledgements}
This research was supported by the UK STFC (F.P.; grant F011083/1), the Israel Science Foundation (T.M.; grant 
655/07) and by NASA (D.F.; Michelson fellowship). The authors thank the anonymous referee for useful comments that helped improve the manuscript.

\bibliography{art2}{}

\end{document}